# Deterministic and Probabilistic P4-Enabled Lightweight In-Band Network Telemetry

Konstantinos Papadopoulos, Panagiotis Papadimitriou, *Senior Member, IEEE*, and Chrysa Papagianni, *Member, IEEE*

*Abstract*—In-band network telemetry (INT), empowered by programmable dataplanes such as P4, comprises a viable approach to network monitoring and telemetry analysis. However, P4-INT as well as other existing frameworks for INT yield a substantial transmission overhead, which grows linearly with the number of hops and the number of telemetry values. To address this issue, we present a deterministic and a probabilistic technique for lightweight INT, termed as DLINT and PLINT, respectively. In particular, DLINT exercises per-flow aggregation by spreading the telemetry values across the packets of a flow. DLINT relies on switch coordination through the use of per-flow telemetry states, maintained within P4 switches. Furthermore, DLINT utilizes Bloom Filters (BF) in order to compress the state lookup tables within P4 switches. On the other hand, PLINT employs a probabilistic approach based on reservoir sampling. PLINT essentially empowers every INT node to insert telemetry values with equal probability within each packet. Our evaluation results corroborate that both proposed techniques alleviate the transmission overhead of P4-INT, while maintaining a high degree of monitoring accuracy. In addition, we perform a comparative evaluation between DLINT and PLINT. DLINT is more effective in conveying path traces to the telemetry server, whereas PLINT detects more promptly path updates exploiting its more efficient INT header space utilization.

## I. Introduction

5G (AND BEYOND) network services raise the need for accurate and high-precision network monitoring in order to promptly detect and reason about application performance degradation related to network faults, link outages, router misconfigurations, and congestion. To this end, network telemetry enables network visibility, facilitating network management

Manuscript received 6 September 2022; revised 11 May 2023 and 24 July 2023; accepted 30 July 2023. Date of publication 3 August 2023; date of current version 12 December 2023. This research was funded by the European Union's Horizon Europe research and innovation program under grant agreement No. 101070487 (NEPHELE). The associate editor coordinating the review of this article and approving it for publication was S. Secci. *(Corresponding author: Panagiotis Papadimitriou.)*

Konstantinos Papadopoulos and Panagiotis Papadimitriou are with the Department of Applied Informatics, University of Macedonia, 540 06 Thessaloniki, Greece (e-mail: konpapad@uom.edu.gr; papadimitriou@uom.edu.gr).

Chrysa Papagianni is with the Department of Informatics Institute, University of Amsterdam, 1098 XH Amsterdam, The Netherlands (e-mail: c.papagianni@uva.nl).

towards the satisfaction of service-level objectives and requirements [1], [2], [3]. Network telemetry typically requires the intervention of the control plane or packet mirroring and forwarding to servers for telemetry analysis [4], [5]. The former approach leads to significant communication overhead for high-precision monitoring, whereas the latter requires additional infrastructure elements that entail scalability limitations.

In-band network telemetry (INT) [6] has emerged as a more viable approach to telemetry analysis, exploiting the recent advances in programmable switching hardware (*i.e.,* P4 [7]). More precisely, INT empowers programmable switches to access and update pre-defined telemetry indicators (*e.g.,* switch ID, buffer occupancy, link utilization) encapsulated into custom packets headers. As such, telemetry data can be directly exported from the dataplane, usually by dedicated telemetry servers at egress points. In this respect, INT can facilitate the correlation of observed application/network performance implications with network bottlenecks, short-lived congestion events, routing misconfigurations, or highly-utilized switches and links [8], [9], [10].

The P4-INT framework encompasses a pre-defined header for the encoding of telemetry metadata into packets. Since INT encodes per-hop information, it inevitably leads to a substantial transmission overhead, which grows linearly with the number of hops. Besides the bandwidth wasted with INT, the extra transmission overhead leads to reduced goodput and excessive flow completion time [11], [12].

To mitigate these problems, we present two lightweight techniques in order to alleviate the transmission overhead of INT. The first technique, namely *Deterministic Lightweight INT* (DLINT), relies on the intuition that telemetry metadata can be spread across multiple packets of a flow, thereby, eliminating the redundancy of telemetry indicators within a flow. Such per-flow aggregation (PFA) is not straightforward though. Instead of any centralized controller intervention which could induce undesirable delays, we opt for switch coordination. DLINT is tailored to path tracing; however, it is extensible to other telemetry applications (*e.g.,* detection of routing misconfigurations) that can benefit from per-flow aggregation. In order to promptly detect potential path updates (which may be associated with transient performance drops), DLINT enables continuous path tracing, which entails additional complexity in its design. In this respect, we utilize a set of telemetry states, maintained within P4 switches, which facilitate switch coordination for the aggregation of telemetry values spread across multiple packets. For telemetry state lookup and update, we utilize Bloom Filters (BF) to cope with the limited amount of memory (*i.e.,* P4 registers) in P4 switches.

We further study an alternative technique for lightweight INT that relies on a probabilistic approach. The so-called *Probabilistic Lightweight INT* (PLINT) empowers each INT node (*i.e.,* P4 switch) to insert telemetry values into the packet with a certain probability. Similar to DLINT, PLINT spreads telemetry values across multiple packets (of a flow), leading to substantially lower transmission overhead. The probabilistic nature of PLINT obviates the need for switch coordination (as opposed to DLINT). However, PLINT raises a requirement of equal probability for inserting telemetry data among all INT nodes (*i.e.,* P4 switches). In order to accomplish this, we rely on reservoir sampling, similar to PINT [11].

This paper extends our previous work [13], which presents a preliminary variant of DLINT with a single telemetry value. The main contributions of this paper are the following:

- We elaborate on a deterministic PFA framework for INT (*i.e.,* DLINT) through per-flow telemetry state update for switch coordination, with the aim of an insignificant and fixed transmission overhead and high monitoring accuracy.
- We present a probabilistic technique for lightweight INT (*i.e.,* PLINT) with a controllable number of telemetry values, based on reservoir sampling. PLINT comprises a simpler lightweight INT approach, without any switch coordination.
- We conduct a comparative evaluation study of both INT techniques using P4 BMv2 software switches [14] in an emulated topology in Mininet.
- We shed light on trade-offs and implications on monitoring accuracy stemming from BF collisions and the probabilistic nature of PLINT. Our evaluation results uncover both gains and limitations due to switch coordination, which can affect the monitoring efficiency under certain conditions.

The remainder of the paper is organized as follows. Section II provides background information on network telemetry and the INT framework. In Section III, we elaborate on the proposed PFA framework and P4 programmable switch coordination for the realization of DLINT. Section IV presents our probabilistic approach to lightweight INT, based on reservoir sampling. In Section V, we discuss our evaluation results. Section VI provides an overview of related work. In Section VII, we discuss potential implications and workarounds with respect to the proposed INT techniques. Finally, Section VIII highlights our conclusions.

## II. IN-BAND NETWORK TELEMETRY

In-band Network Telemetry (INT) is a framework proposed by the P4 Language Consortium [6]. INT allows the collection of network monitoring information from the data plane without the intervention of the control plane. In particular, INT-enabled switches insert an INT header into the packets as they traverse their data path. This header contains instructions for the switch, empowering it to report pre-defined telemetry indicators or insert them in the packet's INT header itself.

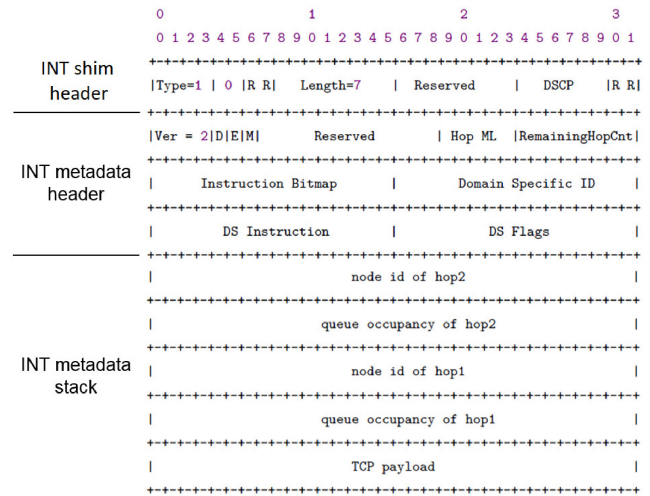

Fig. 1. INT header over TCP in INT-MD mode.

Telemetry variables may include information, such as the switch ID (*e.g.,* in order to trace the path of the packet), the ingress or egress port ID, timestamps, latency, port utilization or queue status and occupancy. Such information can be employed to monitor, debug or (re-)configure the network.

INT can encapsulate the telemetry header into a range of protocol types, such as TCP, UDP, VXLAN, or Geneve [15]. INT can operate on three different modes. In the *Export Data* (XD) mode, the switch is programmed to send telemetry information to the monitoring system (*e.g.,* telemetry server) for the packets that traverse the switch. In this mode, no INT header is inserted to the packets. In the *Embed Instructions* (MX) mode, packets contain an INT header that instructs the switch to report specific information to the telemetry server. Finally, in the *Embed Data* (MD) mode, both instructions and telemetry data are inserted in the packet's INT header.

The INT header depends on the encapsulation protocol and the INT operational mode. More precisely, the INT header consists of an INT metadata header and an INT metadata stack. The former contains general information about the length and the encapsulation of the INT header, whereas the stack includes telemetry data that is inserted by the switches as they forward packets. Fig. 1 illustrates an example of an INT header encapsulated within TCP.

The process of P4's INT for path tracing is shown in Fig. 2. Switches *s1* through *s5* form an INT-enabled part of the network termed as *INT domain*. When a packet enters an INT domain, it is transformed into an *INT packet* by inserting an *INT header*, containing path tracing metadata. In our example, this action is performed by *s1*, called *INT Source*, which encapsulates the INT header and its switch ID. The packet is subsequently forwarded with its standard forwarding process unaffected by INT. The intermediate nodes called *INT nodes* are able to interpret INT packets, and hence, insert their own switch IDs. When the packet reaches *s5*, the switch removes the INT header, sends the metadata stack to the telemetry server, and forwards the packet to its destination.

INT introduces a substantial transmission overhead in packets, which increases linearly with the number of hops, as well as with the number of telemetry values. This transmission overhead in conjunction with the Maximum Transmission Unit (MTU) limitation can severely reduce the packet payload size, with an adverse impact on goodput and flow completion time, as reported in [11]. Our proposed INT techniques alleviate the excessive transmission overhead of INT by aggregating telemetry values across the packets of a flow. This yields significant benefits in telemetry applications, such as path tracing, since it maintains a fixed transmission overhead with a high monitoring accuracy. The functionality of DLINT and PLINT are described in the following sections.

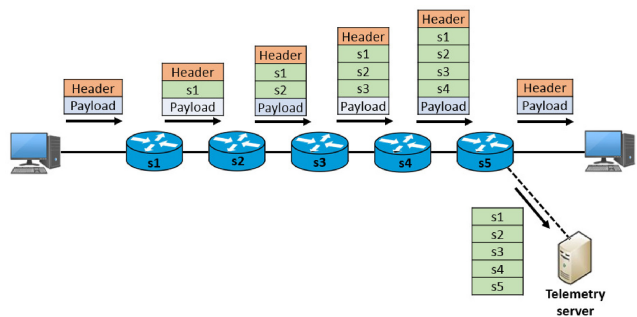

Fig. 2. Example of path tracing with INT.

Fig. 3. Exemplary spreading of telemetry values across several packets repeatedly in order to continuously trace the network path $s1 \rightarrow s2 \rightarrow s3 \rightarrow s4 \rightarrow s5$ in Fig. 2.

## III. DETERMINISTIC LIGHTWEIGHT INT

In this section, we elaborate on our deterministic per-flow aggregation approach. We discuss in detail the interactions between switches in order to correctly implement this aggregation for INT and retrieve the telemetry data with high accuracy. We exemplify our approach with a common telemetry application, i.e., path tracing; however, per-flow aggregation can be employed to confine the transmission overhead for other telemetry applications, as well.

### A. Per-Flow Aggregation

The excessive transmission overhead of INT stems from the fact that switches insert the same telemetry values to each packet (see Fig. 2). In order to alleviate this overhead, we seek to spread the telemetry data across multiple packets within the same flow. In this respect, a path trace, for instance, can be potentially retrieved by inserting a single hop ID within each packet and composing the path from the individual hops that have been stored in the packets. As such, the transmission overhead, e.g., for path tracing, will be bounded to the space required in order to store one hop ID. Such *per-flow aggregation* can turn INT into a more viable approach in terms of goodput performance and flow completion time.

Based on these observations, we have designed and implemented a deterministic lightweight framework for INT (DLINT). The design of DLINT has been dictated by the following two main requirements:

*Correctness:* Telemetry indicators should be correctly retrieved by aggregating telemetry values stored across multiple packets. In terms of path tracing, we should achieve *path completeness*, i.e., ensure that the retrieved path contains all hops with the correct sequence.

*Continuous monitoring:* Telemetry indicators should be retrieved continuously over the entire duration of the flow.

For example, in terms of path tracing, the sequence of hops should be extracted repeatedly, as long as the flow remains active. This enables a telemetry server to promptly detect path changes and correlate this with potential application performance implications that may occur at that time. Such continuous monitoring is depicted in Fig. 3 for a sequence of packets that traverse the path $s1 \rightarrow s2 \rightarrow s3 \rightarrow s4 \rightarrow s5$ of the network in Fig. 2. More precisely, packets 1–5 convey the whole path for a first time. This procedure is repeated twice (i.e., packets 6–10 and packets 11–15), delivering the path trace to the telemetry server three times in total.

Fulfilling these requirements raises the need for coordination among the INT-enabled switches that encapsulate their IDs into the incoming packets for the purpose of path tracing. Switch coordination through a centralized controller would introduce a significant amount of communication overhead and would also contradict with the *in-band* nature of INT. Instead, we opt for a more viable coordination approach, at which each switch maintains per-flow state that indicates the action(s) that it should perform upon each incoming packet. To reduce the amount of state that needs to be maintained by each switch, we leverage on Bloom Filters, which is a data structure widely used for membership lookup, with an adjustable probability of false positives [16]. In the following, we discuss in detail the design of DLINT and the coordination among the switches for path tracing.

### B. Design and Operation

We initially discuss the amount of states that need to be maintained per flow, such that each switch is informed of its turn to insert its hop ID into an incoming packet. At minimum, two states are required, indicating whether a switch has inserted (or not) its hop ID into a packet of a given flow. However, recall the aforementioned requirement for continuous monitoring, which mandates that a flow's path is repeatedly traced over the entire duration of a flow. To this end, we introduce a third state, which facilitates the reset of a switch (telemetry-wise), so that it can insert again its hop ID. Fig. 3 illustrates an exemplary spreading of telemetry values across several packets repeatedly in order to continuously trace the path.

Along these lines, DLINT encompasses the following three telemetry states: (i) *Awaiting Init*, where the switch waits for an INIT signal in order to insert its ID, (ii) *Ready to Insert ID*, where the switch is ready to insert its ID into the following incoming packet of the respective flow, and (iii) *Inserted ID*, where the switch has already inserted its own ID and waits for a signal to revert to the initial state (*Awaiting Init*) in order to re-insert its ID.

We maintain and update 2 bits for these states within each switch using Bloom Filters (BF). BF enables the compression

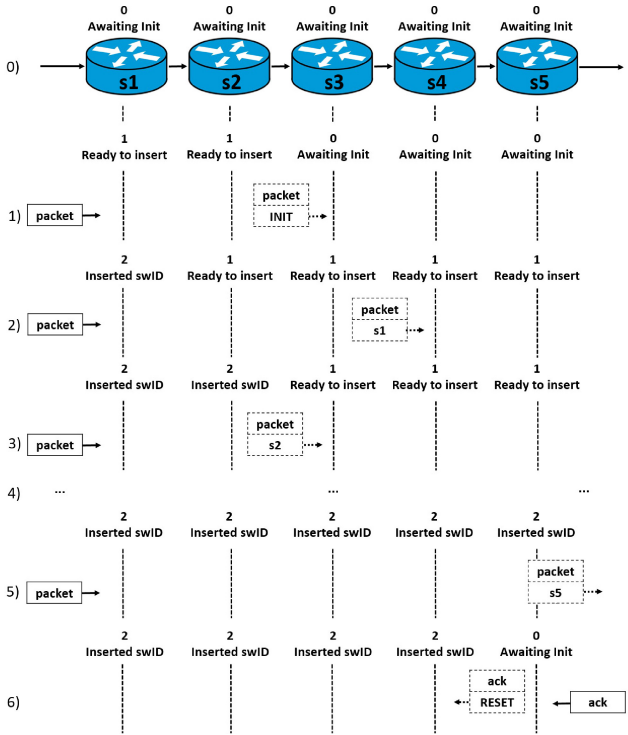

Fig. 4. Sequence of steps taken by DLINT for path tracing across five switches.

of an arbitrary data set into a bit vector and provide membership lookup using hash functions. Assume a BF with size $K$ within each switch. The BF is used in DLINT as follows:

*BF initialization:* The BF is initialized by setting all bits (within each element in the vector) to 0, which designates the initial state (*i.e.,* Awaiting Init).

*BF lookup:* For each incoming packet, the BF computes $m$ hash functions[1] $H_1(), \ldots H_m()$ over pre-defined packet header fields (*e.g.,* 5-tuple) in order to identify the current state for the flow to which the packet belongs, *i.e.,* it checks the value in the position computed by $BF[i = H_v(packet) \bmod K]$, with $v = 1, \ldots, m$.

*BF update:* For any required state update, the BF sets the bits stored in the position retrieved by $BF[i = H_v(packet) \bmod K]$ to the corresponding value (*i.e.,* 0, 1, or 2 for the states *Awaiting Init*, *Ready to Insert ID*, and *Inserted ID*, respectively).

Fig. 4 illustrates the operation of DLINT with five switches. Initially, the BFs in all switches are set to the state *Awaiting Init*. Assume that switch $s1$ receives a packet from a host. The telemetry values are inserted in the following sequence of steps (Fig. 4):

1) The switch $s1$, as the *INT source* for this flow, encapsulates a telemetry header into the packet, containing an *INIT* signal in order to trigger the gradual insertion of telemetry values. As the packet is being forwarded through the switches, each switch receives the *INIT* signal, which triggers the transition to the state *Ready to Insert ID*.

2) When the next packet from the same flow reaches $s1$ (which is in state *Ready to Insert ID*), the switch detects that the packet does not contain any telemetry header. Thus, $s1$ encapsulates a telemetry header into the packet and inserts its own ID. At the same time, the state of $s1$ is changed to *Inserted ID*. The remaining switches along the path observe the encapsulated telemetry header and merely forward the packet without any modification in the telemetry header. The last switch in the path (*i.e.,* $s5$), termed as the *INT sink*, extracts the telemetry header from the packet and forwards the encoded telemetry values to the telemetry server.

3) When the next packet reaches $s1$, the switch forwards the packet without any telemetry value insertion, since $s1$ is in the state *Inserted ID*. Subsequently, the switch $s2$, which is in the state *Ready to Insert ID*, performs the same actions with $s1$ in the previous step, *i.e.,* it inserts its ID within the telemetry header, which is encapsulated in the packet. Next, the state of $s2$ is updated to *Inserted ID*. Similar to the previous step, the switches $s3$ and $s4$ forward the packet, whereas $s5$ conveys the telemetry values to the telemetry server.

4) The actions described in the previous step are repeated for the next two packets, *i.e.,* the switch $s3$ and $s4$ insert their IDs into the first and second packet, respectively.

5) When the next packet is sent over the path, the switches $s1 - s4$ are in the *Inserted ID* state. As such, it is the turn of $s5$ to insert its ID into the packet. Since $s5$ acts as *INT sink*, this obviates the need for inserting a telemetry header into the packet. Instead, $s5$ sends its own ID directly to the telemetry server. This action completes the path trace for the particular flow.

So far, we have explained the steps taken by DLINT for the recording of the path only once. Beyond that point, all five switches in the example of Fig. 4 are in the state *Inserted ID*, meaning that they will not insert any telemetry value to any other packet that belongs to the same flow. In order to perform the path tracing repetitively, all switches should be reset to the initial state (*i.e.,* Awaiting Init). To this end, the *INT sink* emits a *RESET* signal (see step 6 in Fig. 4), which is forwarded along the reverse path, traversing all the switches up to the *INT source*. In DLINT, the *RESET* signal is piggybacked onto TCP ACK packets in order to minimize the communication overhead. Each switch receiving this signal changes its state to *Awaiting Init*. Subsequently, telemetry values are inserted into the packets, following the same steps explained earlier.

Note that although the *RESET* signal resets the state in all switches along the flow's path, resetting the state of only the *INT source* suffices for the correct operation of DLINT. As such, DLINT is not susceptible to path asymmetry, as long as the *RESET* signal reaches the *INT source*.

Consider also that ACK packets (which carry the *RESET* signal) comprise a separate flow that traverses its own (and usually reverse) path. Therefore, path tracing is performed in both directions of the traffic, although it is occasionally interrupted by *RESET* signals that are emitted in the opposite direction and take precedence over switch ID insertion.

---

[1] In our implementation, we use only one hash function.

| Packet # | 6 | 5 | 4 | 3 | 2 | 1 |
|---|---|---|---|---|---|---|
| swID1 | s3 | INIT | s3 | INIT | s3 | INIT |
| swID2 | s4 | s1 | s4 | s1 | s4 | s1 |
| swID3 | s5 | s2 | s5 | s2 | s5 | s2 |

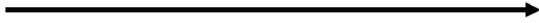

Fig. 5. Telemetry data with three values within packets as they traverse *s5* in the network of Fig. 4.

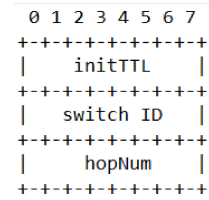

Fig. 6. PLINT telemetry header.

### C. Multiple Telemetry Values

In the previous section, we exemplified the switch coordination for the encapsulation of a single telemetry value within each packet. Since the transmission overhead of a telemetry header with a single value is minimal, we afford to insert multiple telemetry values which can lead to higher monitoring accuracy (from which *e.g.,* time-sensitive flows can benefit).

In this case, first there is a need to reserve space for a predetermined number of telemetry values, so that the telemetry header contains the required number of telemetry value slots. To this end, we require an INT controller, which receives queries from the network operator with respect to the amount of telemetry data. We envisage two ways for the submission of telemetry queries, *i.e.,* a query can explicitly specify the required number of telemetry values or alternatively can indicate a transmission overhead limit, based on which the amount of telemetry values is inferred. Upon the reception of such a query, the INT controller conveys the required number of telemetry values to the P4 switches within the INT domain. As such, the INT source is aware of the telemetry header size (*i.e.,* number of telemetry value slots) that needs to be encapsulated into the packet. Note that the INT controller's role is limited to the identification and communication of telemetry header size; for the coordination among P4 switches, we still rely on the telemetry states, as explained in the previous section.

Although the sequence of steps illustrated in Fig. 4 is, in general, applicable to a telemetry header with multiple values, some additional actions are required on packets. More precisely, the first switch on a path will emit the *INIT* signal, but since there is space in the header for another value, it will also insert its ID in the second field. Hence, the state of the switch immediately evolves from *Awaiting Init* to *Inserted ID*. In the same way that the *INIT* signal is combined with switch ID values, a *RESET* signal can coexist with switch IDs on the reverse path. As such, additional telemetry information can be conveyed through each packet. However, depending on the length of the path, certain telemetry value slots on some packets may be empty. Fig. 5 depicts the telemetry data encapsulated into packets with 3 telemetry values for path tracing along five switches (*i.e.,* similar setup with Fig. 4).

## IV. PROBABILISTIC LIGHTWEIGHT INT

In this section, we elaborate on our probabilistic approach for lightweight INT, namely PLINT. A common element between PLINT and DLINT is the distribution of telemetry values among the packets of a flow, aiming at a low transmission overhead. The main idea behind PLINT is that each P4 switch inserts its own ID with a certain probability. The probabilistic nature of PLINT obviates the need for coordination among the P4 switches within the INT domain. As such, PLINT yields less complexity in its design (compared to DLINT).

Given the uncoordinated nature of PLINT, we should ensure that all switches are in position to encapsulate their ID within each traversing packet with an equal probability. The main difficulty here is that the path length is not known apriori. Thereby, a probability which would be equal among the switches cannot be precomputed and applied.

To address this issue, we resort to reservoir sampling [17], similar to PINT [11]. This algorithm is designed for achieving equal probabilities among a set of entities, the number of which is not known in advance. Reservoir sampling is applied in the case of PLINT, as follows. The first switch in the path always inserts its own ID. Subsequently, the second switch replaces the previous ID with its own, with a probability of 1/2. Likewise, the third switch inserts its own ID with a probability of 1/3, the fourth with 1/4, and so forth. At the end of the path, each switch will end up with an equal probability for the encapsulation of its own ID.

To substantiate this claim, we calculate the corresponding probabilities. For a switch ID to prevail in a packet, it needs to be inserted by the corresponding switch and not to be replaced by any of the downstream INT nodes. The probability that the $i^{th}$ node will insert its ID is $\frac{1}{i}$, whereas the probability that the next node will not replace this ID is $(1 - \frac{1}{i+1}) = \frac{i}{i+1}$. Therefore, the probability of an ID prevailing is:

$$\frac{1}{i} \cdot \frac{i}{i+1} \cdot \frac{i+1}{i+2} \cdot \frac{i+2}{i+3} \cdot \ldots \cdot \frac{n-1}{n}$$

which equals to $\frac{1}{n}$. In essence, by employing reservoir sampling, no switch can prevail in telemetry data encapsulation, although there is no information about the downstream INT nodes. Similar to DLINT, telemetry data is aggregated among multiple packets by the telemetry server, enabling the reconstruction of the complete path. In this respect, we rely on TTL, as we explain later on.

PLINT operations are based on a telemetry header, where the telemetry data can be encapsulated. As shown in Fig. 6, the telemetry header contains the following fields: (i) *initTTL* which stores the initial TTL value of the packet (*i.e.,* the TTL value carried by the packet at its entrance into the INT domain - practically, the TTL seen by the INT source), (ii) the *swID* where the switch ID is stored, and (iii) the *hopNum* used to store the hop sequence by subtracting the current TTL value from *initTTL*.

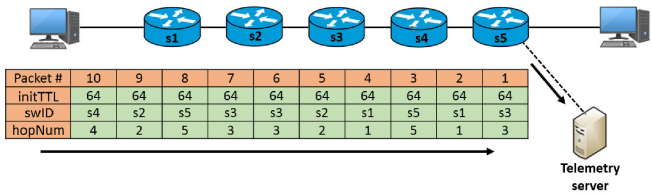

Fig. 7. Telemetry data delivery with PLINT.

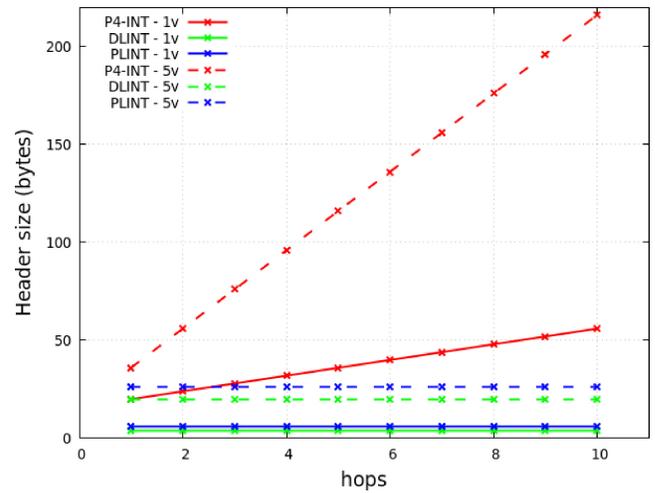

Fig. 8. Transmission overhead of P4-INT, PLINT and DLINT with one and five telemetry values.

Since switch IDs arrive at the telemetry server in a random order, *hopNum* is essential to the telemetry server for the reconstruction of the path. The complete path is reconstructed when all switch IDs of the path have been received. In the example of Fig. 7, ten packets are required for the delivery of the IDs of five switches.

PLINT also supports the encapsulation of multiple telemetry values within a packet. In this case, the telemetry header contains multiple *swID* fields along with corresponding *hopNum* fields. This operation conveys more dense telemetry information to the telemetry server. Similar to DLINT, the INT controller can inform the INT source regarding the amount of required telemetry values per packet, upon the reception of queries from the network operator.

One additional aspect worth noting is the application of reservoir sampling with multiple telemetry values. In this respect, each switch treats the various telemetry values within each packet as independent values. Hence, each telemetry value is replaced based on a probability which is independent from the other values. For instance, assuming a header with three telemetry values, the first switch in the path inserts its ID within all three *swID* fields of each packet. Subsequently, *swIDs* may be replaced based on an independent probability calculated by each node, as explained earlier. As such, a packet may end up with identical values in more than one *swID* fields. This is a side-effect of the uncoordinated and probabilistic nature of PLINT.

## V. EVALUATION

In this section, we evaluate the efficiency and path tracing accuracy of DLINT and PLINT using Mininet [18]. We initially present our evaluation environment (Section V-A) and subsequently, we discuss our evaluation results in terms of transmission overhead (Section V-B), path tracing efficiency (Section V-C), and path update detection (Section V-D).

### A. Evaluation Environment

Both DLINT and PLINT have been implemented using the BMv2 P4 software switch [14]. Our evaluations are conducted on the BTN network topology [19], which consists of 27 nodes and is emulated in Mininet [18]. In particular, we inject traffic using D-ITG [20] into four intersecting paths. The injected traffic encompasses approximately 400 flows, whose size follow a Zipf-like distribution (inline with the distribution of Internet traffic in terms of flow size [21]). The starting time of each flow is picked randomly. The duration of the experiment is 60 seconds.

Packets with encapsulated telemetry headers are collected at the INT sinks using tcpdump [22] and parsed accordingly. We rely on Python for the instrumentation of our tests. The evaluation results presented in the following section are averaged across multiple runs.

### B. Transmission Overhead

We initially perform a comparison between DLINT and P4.org's INT (termed as P4-INT) in terms of transmission overhead. INT uses a varying-size header [6], whose size depends on the number of telemetry values being monitored and the number of hops. Since P4-INT uses 4 bytes for the encapsulation of a telemetry value into its header, we also set the size of a telemetry value to 4 bytes for DLINT (for a fair comparison). Fig. 8 illustrates the transmission overhead for P4-INT compared to PLINT and DLINT variants, with one and five telemetry values. Whereas DLINT and PLINT yield a fixed transmission overhead, P4-INT's overhead grows with the number of hops and the number of values. For instance, with one telemetry value, P4-INT requires 36 bytes in the header after 5 hops, in contrast to DLINT and PLINT which yield a fixed transmission overhead of 4 and 6 bytes, respectively. With five telemetry values, P4-INT utilizes 116 bytes for a five hop path versus 20 and 26 bytes for DLINT and PLINT, respectively. Hence, P4-INT leads to an increasing transmission overhead, which grows further as we insert additional telemetry data, such as switch buffer occupancy, egress ports, etc.

### C. Path Tracing Efficiency

In the following, we compare the efficiency of DLINT and PLINT in terms of conveyed path tracing information. To this end, we measure the number of path traces received by the telemetry server with each method. As explained earlier, each method continuously records the path and conveys the respective indicators to the telemetry server. The experiments are performed with a diverse number of telemetry values per packet, ranging from one to five. With respect to DLINT, we conduct our tests with diverse BF sizes in order to quantify the impact of BF collisions on monitoring efficiency.

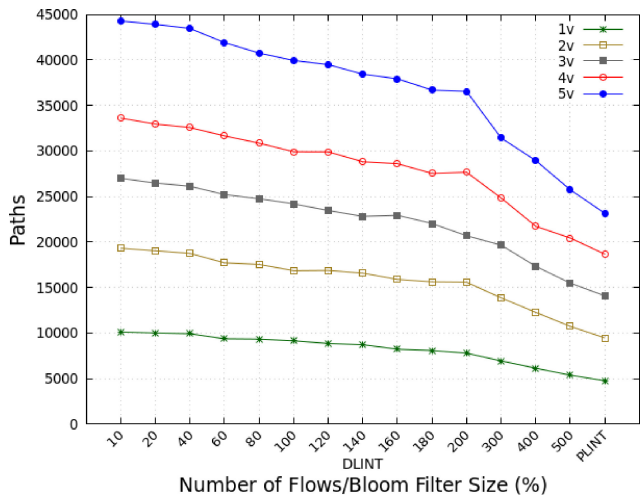

Fig. 9. Number of paths conveyed by DLINT and PLINT with a diverse range of telemetry values. The proportion of flows to BF size applies only to DLINT.

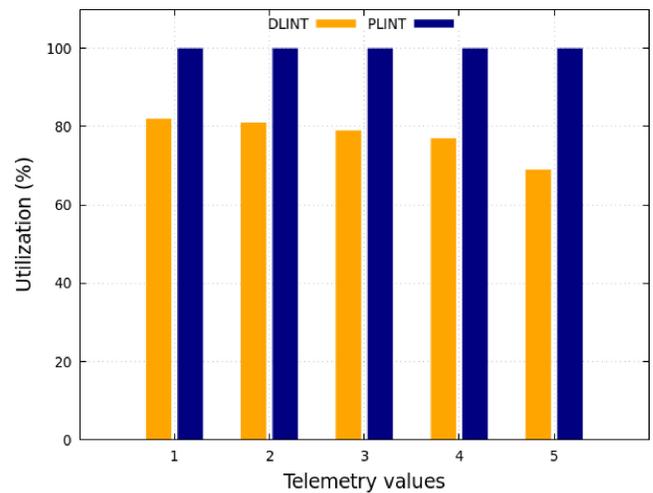

Fig. 10. INT header space utilization with a diverse range of telemetry values.

Fig. 9 illustrates the number of path traces conveyed with DLINT and PLINT across the range of telemetry values. An increasing number of telemetry values obviously yields more dense telemetry information, delivering more fine-grained monitoring data, which essentially corresponds to a larger number of path traces. However, note that two telemetry values do not deliver exactly twice the number of path traces (compared to a single telemetry value), since either some telemetry values may not be used or instead they may be utilized for control messages (*e.g.,* INIT, RESET, etc.). This observation is more prevalent for DLINT, at which the monitoring gain (*i.e.,* number of path traces conveyed) slightly diminishes with a larger number of telemetry values.

According to Fig. 9, BF has a notable impact on the monitoring efficiency of DLINT. As BF size decreases, hash collisions impair path tracing, delivering path traces that are incomplete (which we discard from the set of path traces reported in Fig. 9). When BF equals the number of monitored flows (100% bar), the path traces diminish by up to 10%. The decrease in the number of monitored paths is more severe (*i.e.,* up to 47% with a single telemetry value), when the number of flows is $5x$ the BF size (500% bar).

Although the BF size has a perceptible effect on monitoring accuracy, its impact is not adverse, since a significant number of complete paths is still conveyed to the telemetry server, even with a smaller BF. This essentially stems from the following reasons. First, not all flows collide in every switch; flows may also collide in the same BF slot but not at the same time. Hence, BF collisions are less likely for flows with low degree of temporal correlation. Furthermore, BF collisions may disrupt a few path traces of a flow, whereas the rest of the path traces can be delivered complete.

It is worth noting that although BF collisions can lead to a certain degree of incomplete path traces, they can still contain correct and valuable path trace information. That is, when collisions take place, the path trace will not encompass all switch IDs, but it would instead miss or repeat a subset of switch IDs. However, note that no wrong switch ID will be conveyed. As such, we emphasize on the fact that incomplete path traces can still hold valuable path trace information, which can be utilized by the telemetry server.

To gain more insights into the monitoring efficiency of PLINT and DLINT, we quantify the utilization of INT header space. In this respect, Fig. 10 illustrates that PLINT yields more efficient INT header space utilization (albeit with random switch ID values). The lower efficiency of DLINT in terms of INT header space utilization stems mainly from the following reasons. First, a small subset of the INT header fields are used for coordination signals, such as INIT and RESET. In addition, when all switch IDs of the path are inserted, the leftover fields remain vacant, until the next signal resets the process and re-initiates path tracing. Roughly speaking, tracing a 10-hop path with four telemetry values will leave two of the INT header fields empty in the last packet (without counting the coordination signals). Furthermore, a number of packets traverse without any INT header, since these packets are being forwarded while a P4 switch is waiting for a RESET signal. Our evaluation results indicate that this fraction of packets (*i.e.,* without INT header) is negligible (approximately 0.5% of all packets).

As a result, PLINT conveys more switch IDs, as shown in Fig. 11. In particular, PLINT delivers 5% more switch IDs than DLINT with 1 telemetry value per packet. This margin increases to 31% with five telemetry values. However, the randomness of PLINT outweighs this advantage over DLINT, meaning that the increased number of delivered switch IDs do not lead to equivalent gains in terms of conveyed path traces, as shown in Fig. 9.

In order to better understand this implication, we conduct a deeper investigation of the telemetry indicators conveyed by PLINT. Although PLINT delivers a larger number of switch IDs as the outcome of a better utilization of INT header space (Fig. 10), PLINT may insert duplicate switch IDs within INT headers, since these IDs are picked randomly and independent from the other IDs that have been inserted in the same packet. This problem is exacerbated with a larger number of telemetry values per packet. For instance, with five telemetry values, certain encapsulated switch IDs may be identical, thereby, carrying less telemetry information than a packet with five unique switch IDs.

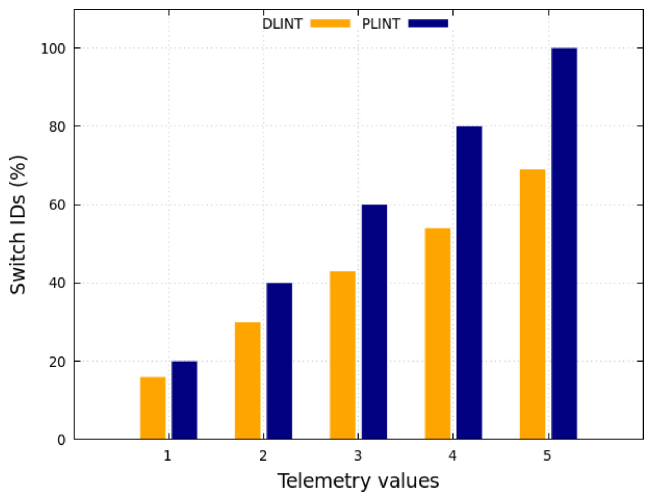

Fig. 11. Percentage of switch IDs conveyed with a diverse range of telemetry values.

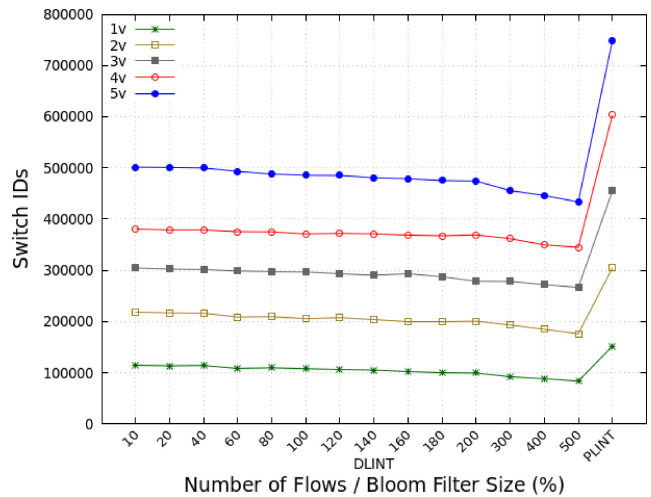

Fig. 13. Number of switch ID delivered by DLINT and PLINT with a diverse range of telemetry values. The proportion of flows to BF size applies only to DLINT.

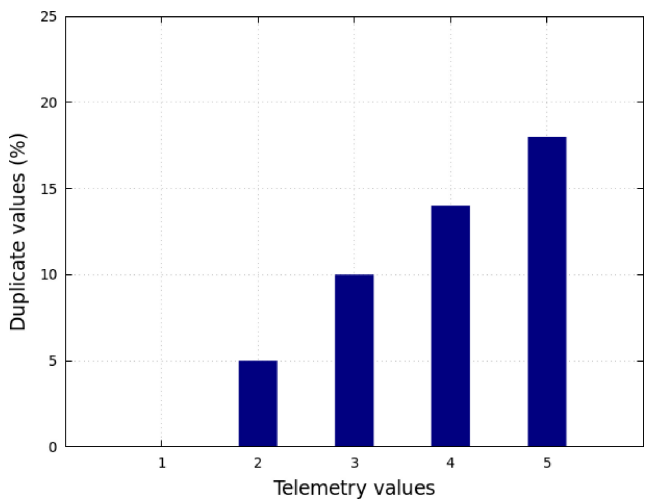

Fig. 12. Percentage of duplicate telemetry data inserted by PLINT with a diverse range of telemetry values.

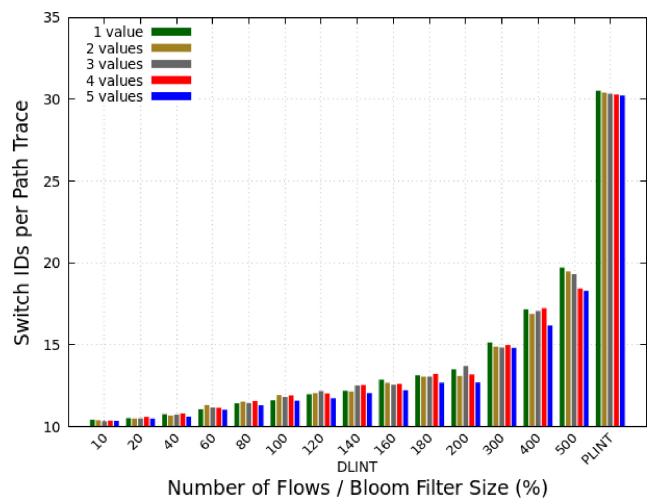

Fig. 14. Average number of switch IDs per path for DLINT and PLINT with a diverse range of telemetry values. The proportion of flows to BF size applies only to DLINT.

In this respect, Fig. 12 illustrates the percentage of duplicate telemetry values encapsulated by PLINT with telemetry fields (per packet) ranging from one to five. This plot essentially corroborates our intuition regarding the trend for duplicate telemetry data versus the number of telemetry values. In particular, we observe a nearly linear increase of duplicate telemetry data, as more telemetry values are encapsulated into the INT header of each packet. In particular, packets with two telemetry values yield a degree of 5% telemetry data redundancy, whereas this reaches 18% in the case of five telemetry values. In essence, this redundancy outweighs the gains of PLINT in terms of INT header space utilization, as reported in Fig. 10. Eventually, the telemetry information conveyed by PLINT is not as dense as implied by Fig. 11. In fact, a subset of switch IDs delivered by PLINT consists of duplicate telemetry values, which do not lead to additional path traces in the telemetry server. This observation, in part, substantiates the discrepancy between the number of conveyed paths (Fig. 9) and the number of delivered switch IDs (Fig. 13), i.e., PLINT conveys fewer path traces to the telemetry server, although it delivers a larger amount of telemetry data (compared to DLINT).

To shed more light into this discrepancy, we measure the number of switch IDs per path trace for both INT methods. The corresponding result is depicted in Fig. 14. In essence, this plot indicates the average number of telemetry values (switch IDs in our case) required at the telemetry server in order to construct a complete path trace. Note that the paths, where the measurements take place, have a hop-count of 10–11. In terms of DLINT, we observe an increasing trend in the number of required switch IDs, as the BF size decreases. This finding is inline with our previous discussion regarding the moderate implications of BF collisions on monitoring accuracy. In principle, as BF collisions increase (*i.e.,* due to a smaller BF), missing or out-of-order switch IDs essentially enforce the delivery of additional telemetry data in order to accurately compose each path.

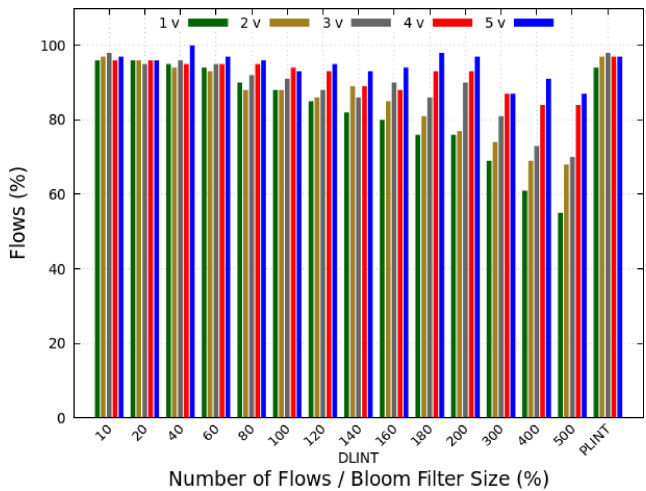

Fig. 15. Percentage of flows with path update detection. The proportion of flows to BF size applies only to DLINT.

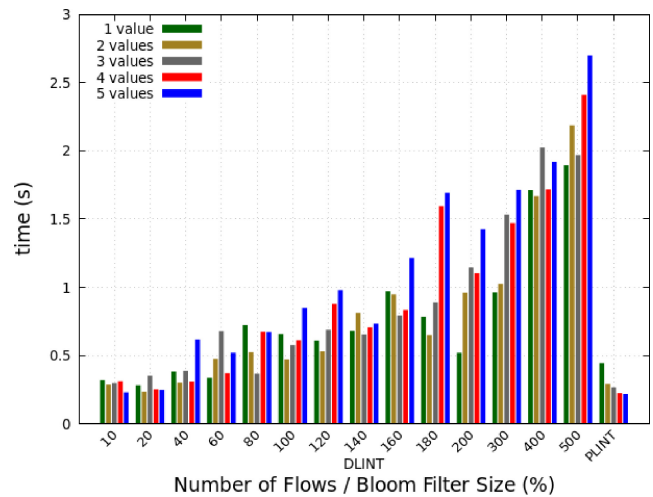

Fig. 16. Time required for path update detection. The proportion of flows to BF size applies only to DLINT.

However, the most interesting and, to some extent, unexpected observation from Fig. 14 is the significant margin between the two INT methods. Whereas DLINT requires 10–20 switch IDs per path (depending on the BF size), PLINT requires the delivery of approximately 30 switch IDs for the path construction at the telemetry server (*i.e.,* $3x$ the path length). As such, even when BF collisions take a toll, DLINT still yields more efficiency, since a smaller number of INT packets are required in order to construct a path. Essentially, Fig. 14 corroborates the tendency of PLINT to deliver random and unstructured telemetry data, as the side-effect of lack of coordination among P4 switches. This issue constitutes the main reason for the aforementioned discrepancy of PLINT between the volume of delivered telemetry data (Fig. 13) and the amount of complete path traces conveyed to the telemetry server (Fig. 9).

### D. Path Update Detection

Furthermore, we evaluate the responsiveness in detecting path updates. In this particular experiment, we enforce an update in all paths and examine whether the path change has been detected, as well as how timely the detection has been. The path update takes place at the $30^{th}$ second of the experiment (recall that the total duration of the experiment is 60 sec).

According to Fig. 15, the path update detection of DLINT is on par with PLINT, as long as the BF is sufficiently large (in relation to the number of flows). However, when BF collisions coincide with the path update, DLINT's detection ability is undermined, and as such, only a fraction of path updates are detected. Our logs indicate that DLINT fails to detect path updates under the following conditions: (i) BF collisions occur during the path update, (ii) the path update takes place during the end of the flow lifetime. This combination of events renders DLINT less responsive to path updates. On the other hand, despite the delivery of a substantially larger number of switch IDs for the construction of a path (Fig. 14), PLINT yields more efficiency in terms of path update detection, since it exhibits a more predictable behavior (*i.e.,* switch IDs, albeit their random order, are always delivered to the telemetry server, augmenting the detection of path updates).

These observations are also valid with respect to how timely the path update detection is (Fig. 16). More precisely, PLINT yields a timely detection, which is at the same level with DLINT, only when the latter is nearly free of BF collisions (*i.e.,* 10-20% bars). DLINT's responsiveness to path updates is degraded when BF collisions take place, as shown in Fig. 16. In essence, BF collisions tend to have a significant impact, as far as the timely detection of path updates is concerned. On the other hand, PLINT's responsiveness is only affected by the larger number of telemetry data (*i.e.,* switch IDs) required for the construction of a path at the telemetry server. As such, an updated path can be detected by the subsequent path trace, as opposed to DLINT where the subsequent path traces may be corrupted (due to BF collisions), thereby, delaying the detection of the path update.

Finally, we perform a comparison of DLINT and PLINT with a prominent probabilistic INT technique, *i.e.,* PINT [11]. Both PLINT and PINT rely on reservoir sampling for the encapsulation of switch IDs into packets. However, PINT has a variable header space budget and is able to use fractions of bytes for storing path trace information in the packets, whereas PLINT encodes the hop count within the INT header along with the switch ID. The latter provides more precise information of the path and augments the detection of path updates, as we explain later on.

To gain more insights into path detection accuracy, we trigger path updates and measure the percentage of path updates detected (Fig. 17), as well as the time required for the detection (Fig. 18). In order to stress on the differences between the three INT methods under comparison, we introduce a different technique for the identification of path updates. Recall that in our previous experiment (Figs. 15 and 16), we relied on whole path traces for the detection of path updates. Hereby, we introduce the so-called *early path update detection*, which obviates the need for the delivery of a whole path trace in order to identify potential updates in the path. Instead, this detection method can identify changes in a path through individual switch ID changes or even through identical switch IDs that correspond to different hop numbers. This is more critical for the two probabilistic techniques under comparison (*i.e.,* PLINT and PINT), which convey switch IDs in a random order. In this respect, a particular strength of *early path update detection* is that it can capture potential gains from the encapsulation of hop count into the telemetry header.

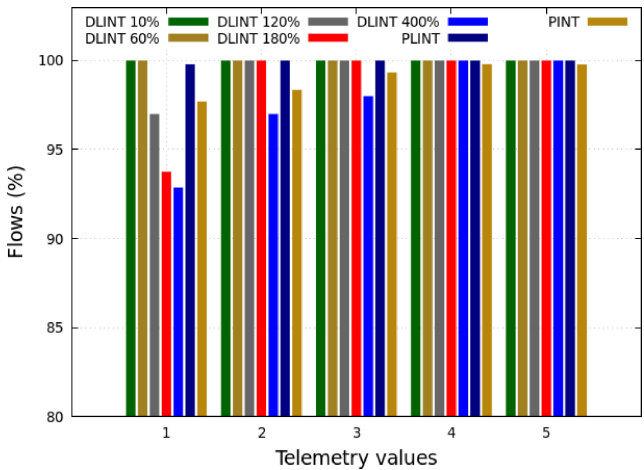

Fig. 17. Percentage of flows with early path update detection. DLINT is associated with diverse proportions of flows to BF size.

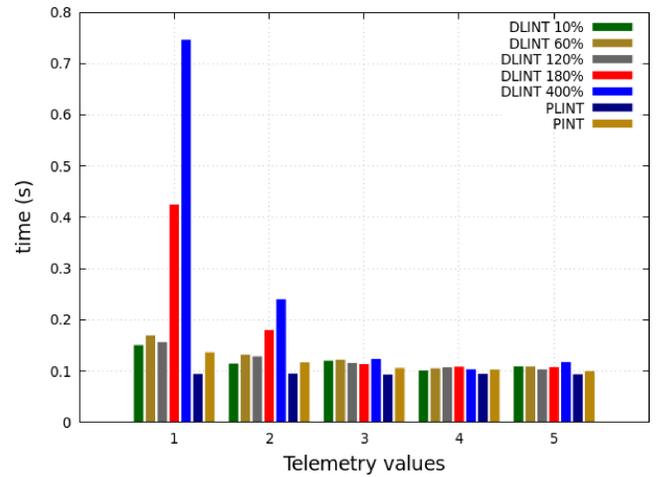

Fig. 18. Time required for early path update detection. DLINT is associated with diverse proportions of flows to BF size.

According to Fig. 17, PINT fails to detect up to 2% of the path updates, as opposed to PLINT that successfully detects nearly all path updates. In terms of detection responsiveness, PLINT detects the path update much faster (up to 40%) than PINT (Fig. 18). Those margins in path update detection efficiency between the two probabilistic INT mechanisms are explained as follows. Hops traced by PINT cannot be readily associated with the initial or the updated path, since PINT does not encapsulate hop numbers into packets (as opposed to PLINT). This inevitably incurs delays in the path update detection. In contrast, the hop numbers (*i.e.,* hop count) conveyed by PLINT facilitate the detection of path updates in the telemetry server. As shown in Figs. 17 and 18, DLINT benefits from multiple telemetry values, which render it more competitive against the probabilistic approaches, with respect to path update detection accuracy and response. Similar to the path update detection based on the whole path trace (*i.e.,* Figs. 15 and 16), DLINT's responsiveness is undermined by BF collisions, whose effect is more prevalent with one and two telemetry values, as well as with a number of flows $\geq 1.8x$ the BF size.

## VI. RELATED WORK

We hereby discuss a range of existing INT techniques, classified into deterministic, probabilistic, and predictive.

### A. Deterministic INT Techniques

The trade-off between monitoring accuracy and the overhead of monitoring data has been recently investigated for INT solutions. Focusing on reducing INT overhead for live network traffic, several approaches adopt a sampling rate for inserting INT fields in packets, *i.e.,* of a flow selectively, thus reducing overhead.

Indicatively, authors in [23] propose a selective INT approach where the ratio for inserting INT headers to the packets of a flow is adjusted, based on the measured difference in value for selected metrics during a predefined period. Simulation results show that sINT can reduce the network overhead compared to standard INT, while authors implicitly document the trade-off in the path update detection time, for different insertion ratios.

PRoML-INT [24] supports INT operations for multi-layer IP-over-optical networks. In order to reduce the corresponding overhead, INT fields are inserted in packets of a flow according to a sampling rate, while each selected packet only carries partial information about the electrical/optical networks on the flow's routing path. Authors show that with a reduced set of INT data samples, their PRoML-INT framework can accurately (100%) identify packet-layer congestion and switch misconfiguration at the dataplane.

Sel-INT [12] adjusts the sampling rate of INT at runtime along with related monitoring information (*i.e.,* locations to collect INT data, INT data types), achieving a trade-off between monitoring accuracy and INT overhead. The sampling rate is set by the SDN controller. Specifically, the controller analyzes historical INT data with the Fourier transform of the traffic trace and sets the sampling frequency as twice the lowest point frequency.

IOAM [25] stands for in-situ operations, management and maintenance. It is a framework deliberated in IETF, for assessing network performance and detecting faults. Metadata is encapsulated within each packet in order to verify its path trace. IOAM can also incorporate timestamps, hop latency, buffer occupancy, TCP sequence numbers and other telemetry data, utilized in network debugging. IOAM is implemented in Cisco Vector Packet Processing (VPP) and Linux. However, IOAM suffers from the same transmission overhead limitations and MTU exceedance of P4-INT, since the INT header increases after each hop.

Alternate marking-performance measurement (AM-PM) [26] is designed to measure packet loss, delay, and jitter in network traffic. AM-PM uses a method to mark specific packets and network devices send telemetry data to the server when the marks are encountered. As such, telemetry measurements are being held between these marked points in a flow. The marking uses minimum packet overhead but processing is required in the switches, since counters and match action tables need to be maintained. AM-PM constitutes a hybrid solution that employs both in- and out-of-band telemetry.

FS-INT [27] supports two sampling strategies; a rate-based one, where telemetry data is inserted into the INT header every $r$ packets and an event-based one where INT transit devices decide upon inserting telemetry data based on some criterion (*e.g.,* queue length over a predefined threshold). The latter yields encouraging results with regards to the accuracy of the obtained measurements compared to the original INT approach, reducing protocol overhead by approximately 50%.

FINT [28], based on a triple bitmap, enables setting telemetry tasks and parameters (*i.e.,* the combination of telemetry metadata types, the INT period, etc.) dynamically at runtime. Flexible INT adaptation without re-deployment reduces the impact of telemetry on network performance, through the dynamic selection of telemetry metadata. To this end, authors propose corresponding greedy and random telemetry metadata selection algorithms. Experimental results show that FINT is effectively reducing the average flow completion time of flows carrying INT data in the network without increasing significantly bandwidth consumption.

A number of approaches have been proposed to ensure full coverage for the network and scalable telemetry. Graph Partitioned INT (GPINT) [29] proposes a heuristic based on Kernighan-Lin's graph partitioning algorithm to find balanced paths in order to forward in-band telemetry information (probes). Apart from minimizing network overhead, the goal is to guarantee the freshness of telemetry information and minimize redundant information. The approach is further enhanced in [29], which provides seamless recovery from link failures and potential INT probe losses by integrating shared queue ring as a reliability feature.

Focusing on the timeliness of telemetry data and the MTU limit, C-INT [30] proposes a clustering approach to divide the topology into several clusters, along with the corresponding data processing pipeline. The cluster header sends the telemetry report per cluster to a (distributed) collector.

Sketchint [31] is an example of a combined INT and sketch-based approach. While INT provides per-flow per-switch measurements at the cost of high network overhead, sketching solutions achieve low network overhead sacrificing accuracy for per-flow measurements. SketchINT collects per-packet INT information, aggregates them by encoding the per-flow information into compact sketches, and periodically reports the sketch to the monitoring system. As such, the bandwidth usage is reduced.

In contrast to most deterministic approaches, our proposed deterministic approach utilizes telemetry states, maintained within Bloom Filters, in order to enable inter-switch coordination for the spreading of telemetry data across the packets of a flow, thereby, leading to a substantial reduction of the transmission overhead.

### B. Probabilistic INT Techniques

PINT [11] uses probabilistic techniques to avoid the hassle of coordination among switches, while it bounds the per-packet overhead. Each switch inserts its own telemetry data with an equal probability to the others. Hence, node IDs reach the INT sink in a random order. When an adequate number of packets is received, the full path of the flow is composed. PINT also uses distributed coding to further break the telemetry data and reduce the necessary space required in each packet. PLINT works for aggregate data and network flows that are not short-lived. Similar to PLINT, PINT exhibit limitation due to its probabilistic approach. More precisely, in PINT, switches randomly decide on adding INT data. Thus, redundant telemetry information can be reported, as opposed to our proposed deterministic approach (*i.e.,* DLINT) that does not yield such redundancy.

INT-label [32] employs a probabilistic labelling algorithm. The algorithm labels device states onto packets based on the number of already collected INT metadata in each packet. A feedback mechanism is used for adapting the label frequency in order to avoid telemetry resolution degradation, due to loss of labelled packets. Experiments show that the probabilistic labelling approach reduces the number of uploaded INT packets by 35.2%, while by employing adaptive labelling, measurement coverage reaches 92% even if 60% of the packets are lost.

Karaagac et al. [33] propose an adaptation of INT for wireless industrial sensors. Telemetry data is inserted in the Information Elements (IE) field of the LR-WPAN MAC frame. Various fields adjust the amount and density of the telemetry indicators that will be conveyed. Different modes of operation instruct intermediate nodes to insert telemetry information either always or based on a probability. Nodes can spread telemetry data on several packets. There is also an option for a node to encapsulate custom telemetry data.

LightGuardian [34] combines in-band telemetry with device-local sketches; it captures per-flow statistics with sketches on programmable switches. The switches periodically split their sketches into sketchlets (sketch fragments) and send them to local or global analyzers by packet piggybacking. The packets that carry a sketchlet are selected with a fixed probability.

### C. Predictive INT Techniques

Recent studies adopt predictive techniques for identifying and filtering redundant or less interesting observations of telemetry data in the data plane.

Indicatively, INT-Filter [4] advocates the use of the same prediction mechanism at the controller and data plane to predict device internal states based on historical data.

Depending on the accuracy of the prediction at the data plane, it notifies the controller employing a single-bit flag to use the predicted value or alternatively inserts the INT data. Authors in [4] also propose additional methods (polynomial fitting) to improve the accuracy of prediction.

In a similar fashion, using LINT [35] each data-plane device estimates the prediction error at the collector, employing the same prediction function as the collector, and depending on this value it reports or skips the current observation. The approach is also applied on a per-flow basis. Such approaches are oblivious to flow duration; however, the overhead reduction in this case depends heavily upon (i) the accuracy of the prediction method and (ii) the stability/predictability of the devices' state.

## VII. Discussion

We hereby discuss some potential implications in the operation of DLINT and PLINT, as well as ways to mitigate these problems.

*BF collisions:* The main limitation of DLINT stems from BF collisions, which tend to occur when the number of monitored flows is larger in comparison with the BF size. At the event of BF collision, incoming packets may interfere with other flows, erroneously modifying their state and eventually leading to undesirable implications for path tracing. An observed implication of a BF collision is that certain hops are missing from the recorded path, since the state in the corresponding BF position may have been updated to *Inserted ID* before any insertion has taken place for that flow. Nevertheless, the continuous path tracing of DLINT can alleviate this issue, since the path can be potentially retrieved from subsequent packets, *e.g.,* when the colliding flow has terminated and, thereby, no interference occurs. Since the size of the BF will be restricted due to the limited amount of memory in P4 switches (and particularly in P4 switch registers, where the telemetry state is expected to be stored), there are certain workarounds for reducing the collisions. One option is to use multiple hash functions in the BF. In theory, this would be beneficial only when flows occupy up to 35% of the BF size. This stems from the equation ($ln2$) $K/N$, which gives the optimal number of hash functions depending on the BF size ($K$) and the number of elements ($N$) stored in the BF [16]. Furthermore, any potential benefits would be outweighed by the additional processing overhead on P4 switches (*i.e.,* due to the computation of multiple hash functions).

Beyond the BF, there are certain ways to reduce the amount of flows that have to be kept in the BF, *e.g.,* monitoring only a subset of the flows. The packets of these flows could be marked (*e.g.,* using IP header fields, such as ToS or flow ID), so that the switch is aware of which packets require telemetry values. Furthermore, additional functionality at the telemetry server or even the control plane can facilitate the resolution of collisions. Alternatively, the telemetry server can rely on consistent path traces, discarding any sporadic traces that deviate from the path monitored over a longer period. Techniques for the resolution of BF collisions will be investigated in future work.

*Short flows:* Since DLINT requires a certain number of packets in order to record the path, it will not be able to trace the path of short flows, *i.e.,* in particular, flows that encompass fewer packets than the number of the hops that they traverse. In case monitoring of short flows becomes critical, we could apply per-flow aggregation beyond a pre-defined number of packets, allowing the encoding of all telemetry values (*e.g.,* hop IDs) within each one of the first packets (*i.e.,* similar to P4-INT), thereby, enabling path tracing (or other telemetry applications) for short flows. To this end, an additional telemetry state could be maintained to designate when to switch from P4-INT to DLINT. In terms of BF collisions, their impact (from maintaining telemetry states of short flows) on other flows will be low, since short flows will yield a small degree of temporal correlation with the majority of the flows being monitored.

*Path updates:* When a flow encounters a path change, packets may traverse switches that have not been initialized for that flow. As such, these switches will not insert any telemetry value to the packet, and as a consequence, these hops will be missing from the path trace. Nevertheless, this problem will be mitigated in the following repetition of the path trace, since the *RESET* signal emitted by the INT sink will be received by all switches in the updated path. Therefore, the path will be recorded correctly in the following repetitions.

*Lost signals:* DLINT relies on signals for certain telemetry state updates. The loss of a signal may leave a switch in an unexpected state and may even stall the INT process for a flow. This problem can be mitigated as follows. If an *INIT* signal is lost, the following INT packet will be unexpected since the switches will be in *Awaiting INIT* state. In this case, if a switch in the *Awaiting Init* receives an INT packet with an encapsulated switch ID, it will assume that *INIT* has been lost and, thereby, will evolve into the *Ready to Insert ID* state. On the other hand, the loss of a *RESET* signal can be more severe, since no other signals follow to tip the switches for this unexpected condition. Consequently, a lost *RESET* signal would stall the process for that flow. This can be remedied by the periodic emission of a PROBE signal for the sole reason of checking the integrity of the telemetry states.

*Duplicate values:* The most notable issue of PLINT is the insertion of duplicate values in the telemetry fields of the INT header of a packet. This problem stems from the probabilistic nature of PLINT and its intensity depends on the number of telemetry fields (*i.e.,* duplicate values are expected to increase with more telemetry fields in the INT header). This shortcoming of PLINT can be partially mitigated in the last hop of the path, where the insertion of telemetry indicators is being finalized (based on the reservoir sampling approach). More precisely, the last-hop P4 switch can inspect the INT header and, in the case of duplicate values, it can insert its own switch ID in order to reduce or even eliminate, if possible,[2] the telemetry data redundancy in the INT header.

---

[2]Not all duplicates can be eliminated, since the switch can only substitute one duplicate value with its own switch ID.

## VIII. Conclusion

In this paper, we elaborated on a deterministic and a probabilistic approach towards lightweight INT, with the aim of confining the transmission overhead while maintaining a high degree of monitoring accuracy. Both approaches perform per-flow aggregation in order to spread the telemetry data across the packets of a flow, and can support an arbitrary number of telemetry values. DLINT requires a form of switch coordination, which is achieved through telemetry states maintained within P4 switches in the INT domain. On the other hand, PLINT operates without any coordination, by empowering switches to encapsulate telemetry values into packets with a certain probability. Employing reservoir sampling guarantees an equal probability among all INT nodes.

Our evaluation results corroborate the feasibility of both lightweight INT approaches. DLINT is resilient to a wide range of BF collision rates, allowing the correct tracing of paths. The comparison between DLINT and PLINT uncovers various trade-offs that affect the efficiency of both INT techniques. More specifically, DLINT requires a smaller number of packets in order to convey path traces to the telemetry server. On the other hand, PLINT yields a more efficient utilization of the INT header space, despite the occasional encapsulation of duplicate switch IDs into the header. PLINT also manages to encapsulate an INT header to all incoming packets, as opposed to DLINT that misses a small fraction of packets when the switches are waiting for the *RESET* signal (*i.e.,* between repetitions of path traces). These advantages of PLINT outweigh (to a certain extent) its main shortcoming, *i.e.,* the larger number of packets required for the collection of monitoring information, due its probabilistic nature. In principle, DLINT is deemed more efficient for capturing path traces; however, this gain diminishes as the proportion of flows over the BF size increases. On the other hand, PLINT outperforms DLINT in terms of path update detection, empowering the telemetry server to detect more promptly and accurately potential path changes of monitored flows.

Future work will be focused on the porting of our INT implementations into other P4 targets, such as P4 switches and SmartNICs. This will empower us to study performance aspects of our INT framework in an experimental environment, such as the delays incurred during telemetry state lookup and the telemetry header encapsulation and decapsulation.

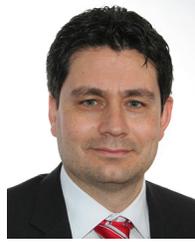

**Panagiotis Papadimitriou** (Senior Member, IEEE) received the B.Sc. degree in computer science from the University of Crete, Greece, in 2000, the M.Sc. degree in information technology from the University of Nottingham, U.K., in 2001, and the Ph.D. degree in electrical and computer engineering from the Democritus University of Thrace, Greece, in 2008. He is an Associate Professor with the Department of Applied Informatics, University of Macedonia, Greece. Before that, he was an Assistant Professor with the Communications Technology Institute, Leibniz Universität Hannover, Germany, and a member of L3S Research Center, Hanover. He has been a (co-)PI in several EU-funded (e.g., NEPHELE, T-NOVA, CONFINE, NECOS) and nationally funded projects (e.g., G-Lab VirtuRAMA, MESON). His research activities include (next-generation) Internet architectures, network processing, programmable dataplanes, time-sensitive networking, and edge computing. He was a recipient of Best Paper Awards at IFIP WWIC 2012, IFIP WWIC 2016, and the Runner-Up Poster Award at ACM SIGCOMM 2009. He has co-chaired several international conferences and workshops, such as IFIP/IEEE CNSM 2022, IFIP/IEEE Networking TENSOR 2020–2023, IEEE NetSoft S4SI 2020, IEEE CNSM SR+SFC 2018–2019, IFIP WWIC 2016–2017, and INFOCOM SWFAN 2016. He is also an Associate Editor of IEEE TRANSACTIONS ON NETWORK AND SERVICE MANAGEMENT.

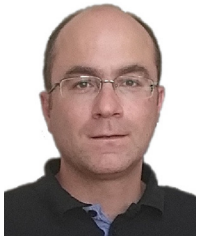

**Konstantinos Papadopoulos** received the B.Sc. degree in computer science from the Aristotle University of Thessaloniki, and the M.Sc. degree in information and communication technology from International Hellenic University. He is an Assistant Researcher with the Department of Applied Informatics, University of Macedonia, Greece. His research interests include programmable networks, network telemetry, software-defined networking, and cloud technologies.

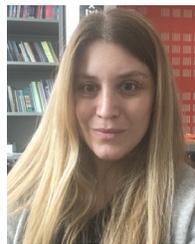

**Chrysa Papagianni** (Member, IEEE) is currently an Assistant Professor with the Informatics Institute, University of Amsterdam. She is part of the Multiscale Networked Systems Group that focuses its research on network programmability and data-centric automation. Prior to joining UvA, she was a Network Research Engineer with Bell Labs Antwerp, as part of the Network Service Automation Laboratory. From 2016 to 2018, she was a Research Scientist with the Institute for Systems Research, University of Maryland, USA. Her research interests include the area of programmable networks with emphasis on network optimization and the use of machine learning in networking. She has participated in various EU FIRE and 5G-PPP Projects, such as Fed4FIRE, OpenLab, NOVI, and 5Growth working on issues related to network slicing.